\documentclass[twoside, epsfig]{article}

\input oejv.sty

\usepackage{multirow}

\begin{document}

\OEJVhead{December 2020}
\OEJVtitle{CK Aqr time keeping. Evidence for a third body}
\OEJVauth{Bonnardeau, Michel$^1$ }
\OEJVinst{MBCAA Observatory, Le Pavillon, 38930 Lalley, France, {\tt \href{mailto:arzelier1@free.fr}{arzelier1@free.fr}}}

\OEJVabstract{Photometric measurements of the contact binary system CK Aqr are presented. 49 new times of minimum were obtained between 2005 and 2020. Along with already published observations, this allows the derivation of an improved ephemeris. The resulting O-C diagram shows oscillations which are interpreted as the light time travel effect due to a third component, with a period of 8.2 years.
}

\begintext

\section{Introduction}\label{secintro}
CK Aquarii (RA=21h 01min 02.3s DEC=-11$^{\circ}$ 04' 27" (2000.0)) was recognized as a contact binary star (W UMa type) by \citet{leborgne89}. According to the GCVS, its magnitude varies between 12.86 and 13.47, with a period of 0.2833718 day. 
\\ \\
In this paper, I present 13 seasons of photometric observations to obtain light curves, spanning from 2005 to 2020. 
\section{Observations}
\begin{figure}[htbp]
	\centering
    \includegraphics[width=14cm]{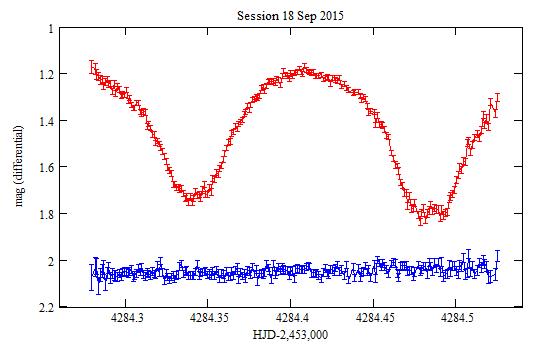} 
	\caption{Red: the differential magnitudes (C filter) for CK Aqr, Blue: for the check star GSC 5774-1263. The error bars are $\pm$ the 1-sigma statistical uncertainties (quadratic sum of the statistical uncertainties for the variable and for the comparison). The two eclipses for one orbit are visible, at $4284.341 \pm 0.001$ and $4284.484 \pm 0.002$ THJD.}
	\label{fig1}
\end{figure}

I observed CK Aqr with a 203 mm f/6.3 Schmidt-Cassegrain telescope, either a Johnson V filter or a clear (C) filter and a camera with a KAF401E CCD (mostly red sensitive). I made time series with individual exposures of 200 s with the V filter and 60 s with the clear one. For the differential photometry, I use GSC 5774-1024 as a comparison star. The magnitudes and colors of the comparison star and of the variable star may be estimated from the CMC14 catalog, using the transformation of \citet{bilir08} and \citet{smith02} as $V\approx11.81$, $B-V\approx0.98$ for the comparison star, 
$B-V\approx0.87$ for CK Aqr. Therefore the two stars have roughly the same color, with the comparison being a bit brighter and at 3' from the variable. An example of a light curve is shown in Fig.~\ref{fig1}.
\\ \\
I obtained 1187 V filter images, 3702 clear filter images, making 43 light curves. The photometric measurements are available as the machine-readable file \textit{photometry.dat} in the appendix. The bottom of each eclipse shows "bumps" and is variable from one eclipse to the other. The eclipse minimum timing is then done by eye, with an uncertainty between 1.5 min and 3 min, depending upon the shape of the light curve. This allows the determination of 49 times of minimum (ToM), listed in Tab.~\ref{tab1} and Tab.~\ref{tab2}.

\begin{table}[htbp]
	\caption{List of the observed times of minimum.}\vspace{3mm}  
	\centering
	\begin{tabular}{p{15mm} p{25mm} p{20mm} c p{10mm} r}
		\hline
	 \multirow{2}{*}{Season} & ToM  & \multirow{2}{*}{uncertainty} & \multirow{2}{*}{Session} & \multirow{2}{*}{filter} & number of   \\
	 & [HJD-2,453,000] & & & & images \\ \hline \hline
	\multirow{2}{*}{2005} 	& 571.5995	& 0.0015  & 19 Jul  & C  & 182 \\
		 	& 653.352 	& 0.0015  & 9 Oct  & C  & 27 \\		  
		 	\hline
		\multirow{2}{*}{2009} 	& 2061.433	& 0.002  & \multirow{2}{*}{17 Aug}  &\multirow{2}{*}{V} & \multirow{2}{*}{96} \\ 
		 	& 2061.576 	&  0.002 &  &  & \\  
		\hline	
		\multirow{8}{*}{2010} & 2391.563 &	0.0015 &	13 Jul& 	V	& 36\\ 
	&	2393.546 &	0.001 &	15 Jul&	V&49 \\
	&	2396.522 &	0.001 &	18 Jul& 	V&63 \\
	&	2426.562 &	0.001 &	17 Aug& 	V&49 \\
	&	2440.4475 &	0.0015 &	31 Aug& 	V&69 \\
	&	2441.439 &	0.001 &	1 Sep &	V& 76\\
	&	2443.422& 	0.0015 &	3 Sep &	V&76 \\
	&	2498.254 &	0.001 &	28 Oct& 	V&65 \\ 
		\hline
	\multirow{5}{*}{2011} & 	2740.535 &	0.001& 	27 Jun& 	V &	55\\
	&	2803.444 &	0.0015 &	29 Aug &	V &	79 \\
	&	2833.340 &	0.0015 &	28 Sep &	V &	68 \\
	&	2835.324 &	0.0015& 	30 Sep &	V &	47 \\
	&	2836.460 &	0.001 &	1 Oct &	V 	& 60\\
		\hline
	\end{tabular}\label{tab1}
\end{table}

\begin{table}[htbp]
\caption{continued from Table 1.}\vspace{3mm}  
\centering
\begin{tabular}{p{15mm} p{25mm} p{20mm} c p{10mm} r}
	\hline
		\multirow{5}{*}{2012} &	3135.555 &	0.0015 &	26 Jul &	V &	64 \\
	&	3148.448 &	0.0015&	8 Aug &	V &	62 \\
	&	3158.365 &	0.002 &	18 Aug &	V &	53 \\
	&	3166.440 &	0.001 &	26 Aug &	V &	72 \\
	&	3252.301 &	0.0015 &	20 Nov &	V &	48 \\
		\hline
	\multirow{6}{*}{2013} &	3522.496 &	0.0015& 	17 Aug& 	C &	218 \\
	&	3539.356 &	0.0015 &	\multirow{2}{*}{3 Sep} & \multirow{2}{*}{C} &  \multirow{2}{*}{203} \\
	&	3539.500 &	0.001&&&\\
	&	3558.3435 &	0.0015 	&22 Sep &	C& 	104\\
&	3637.264 &	0.0015 &	10 Dec& 	C& 	90 \\
&	3638.252 &	0.002 &	11 Dec& 	C &	73 \\
		\hline
		\multirow{2}{*}{2014}&	3829.529 &	0.0015 &	20 Jun &	C &	141 \\
	&	3887.479 &	0.001 &	17 Aug &	C &	186 \\
		\hline
	\multirow{2}{*}{2015}&	4284.341 &	0.001 &	\multirow{2}{*}{18 Sep}&\multirow{2}{*}{C}&\multirow{2}{*}{221} \\
	&	4284.484 &	0.002&&&\\
		\hline
	\multirow{3}{*}{2016}&	4614.472 &	0.001 &	13 Aug &	C &	170 \\
	&	4640.401 &	0.001 &	8 Sep& 	C&	137 \\
	&	4730.231 &	0.0025& 	7 Dec &	C &	85 \\
		\hline
	\multirow{3}{*}{2017}&	4930.577 &	0.0015& 	25 Jun &	C& 	141 \\
	&	4987.3935 &	0.001 &	\multirow{2}{*}{21 Aug} & 	\multirow{2}{*}{C} & \multirow{2}{*}{261} \\
	&	4987.535 &	0.001&&& \\
		\hline
	\multirow{2}{*}{2018}&	5296.555 &	0.0015 &	26 Jun &	C &	149 \\
	&	5389.3585 &	0.001 &	27 Sep &	C& 	204 \\
		\hline
		2019 &	5721.4695 &	0.001 &	25 Aug &	C &	202 \\
		\hline
		\multirow{8}{*}{2020}&	6096.3665 &	0.001 &	\multirow{2}{*}{3 Sep}&	\multirow{2}{*}{C}&	\multirow{2}{*}{221} \\
		&	6096.508 &	0.001&&& \\
		& 6123.285 & 0.0015  &	\multirow{2}{*}{30 Sep}&	\multirow{2}{*}{C}&	\multirow{2}{*}{213} \\
		& 6123.427 & 0.002&&& \\
		& 6153.3235 & 0.002 & 30 Oct & C & 162 \\
		& 6172.31 & 0.0015 & 18 Nov & C & 117 \\
		& 6176.275 & 0.002 & 22 Nov & C & 109 \\
		& 6181.233 & 0.001 & 27 Nov & C & 86 \\
		\hline
	\end{tabular}\label{tab2}
\end{table}

\section{Ephemeris for the ToMs}
\begin{figure}[htbp]
	\centering
	\includegraphics[width=12.5cm]{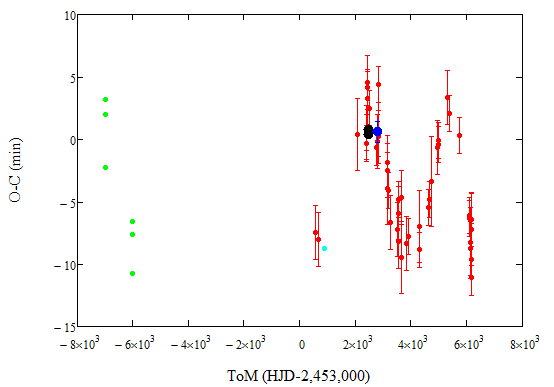}
	\caption{Green: \citet{leborgne89}; Black: \citet{hubscher11}; Blue: \citet{banfi12}; Cyan: GCVS; Red: my observations.}
	\label{fig2}
\end{figure}

So I observed 49 ToMs from 2005 to 2020. To derive an improved ephemeris for the eclipses, I also used published observations:
\\
6 ToMs from \citet{leborgne89}, from 1984 to 1987;
\\
2 ToMs from \citet{hubscher11}, in 2010;
\\
1 ToM from \citet{banfi12}, in 2011;
\\
1 ToM in 2006 from the ephemeris in the General Catalog of Variable Stars (GCVS), 2013 (in JD, the heliocentric correction is 259.5 s).
\\ \\
This is a total of 59 ToMs. They are converted in BJD and are listed in the machine-readable file \textit{ToM.dat} in the appendix, along with the eclipse numbers computed from the GCVS ephemeris.
\\ \\
These 59 ToMs are fitted with the linear ephemeris $ToM=T_1+P_1.N$ where $N$ is the eclipse number, using a least squares calculation weighted with the uncertainties on the ToMs (for the ToMs from \citet{leborgne89} and the one from the GCVS I assume an uncertainty of 0.001 day). The result is:
\\ \\
$T_1=2453892.80782(28)$ BJD
\\
$P_1=0.28337219679(12)$ d 
\\ \\
The resulting O-C diagram is shown  Fig.~\ref{fig2}.
\\ 
I also tried to fit the ToMs with a quadratic ephemeris but this did not yield a significant derivative of $P_1$.
\section{Interpretation of the O-C diagram}
\begin{figure}[htbp]
	\centering
	\includegraphics[width=14cm]{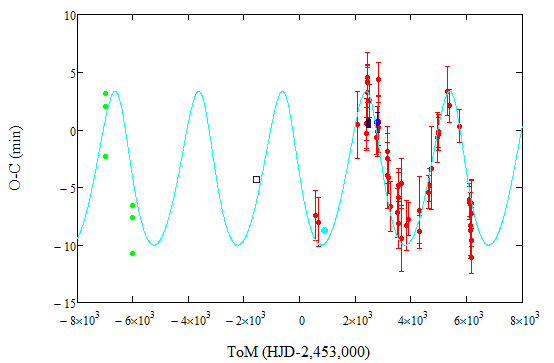}
	\caption{Green: \citet{leborgne89}; Black: \citet{hubscher11}; Blue: \citet{banfi12}; Cyan: GCVS; Red: my observations, Cyan line: the fit with the LTTE, Black square: computed O-C from the 1999 ROTSE observations. }
	\label{fig3}
\end{figure}

The O-C diagram of Fig.~\ref{fig2} shows what appears to be oscillations. I interpret this as a light time travel effect (LTTE) due to a third body with a period around 8.5 yr. The LTTE is given by 
\\ \\
$LTTE(t)=\displaystyle\frac{-r(t)\sin(i)}{c}\sin(\phi(t)+\omega)+\frac{ae\sin(i)}{c}\sin(\omega)$
\\ \\
with the true anomaly $\phi(t)$ given by:
\\ \\
$\phi(t)=2\arctan${\LARGE$[$}$\displaystyle\sqrt\frac{1+e}{1-e}\tan$
{\Large$[$}$\pi\displaystyle\frac{t-t_0}{P}+\frac{\sqrt{1-e^2}}{2(1+e\cos(\phi(t)))}e\sin(\phi(t)${\Large$]$}{\LARGE$]$}
\\ \\
and
\\ \\
$r(t)=\displaystyle\frac{a(1-e^2)}{1+e\cos(\phi(t))}$
\\
$t$ the time
\\$P$ the period
\\$t_0$ the time of passage at the periastron
\\$a$ the semi-major axis
\\$\omega$ the periastron longitude (from the node line)
\\$e$ the eccentricity
\\$i$ the inclination.
\\ \\
The O-C diagram is fitted with the $LTTE(t)$ function using a Monte Carlo method. I test randomly selected parameters and I retain those that give the smallest residuals. I make 10 runs of 1 million trials each. The resulting averages and standard deviations from the runs are: 
\\ \\
$e=0.273 \pm 0.014$
\\$a\sin(i)=0.8078 \pm 0.0051$ AU
\\$\omega=-65.75 \pm 8.7$  $^{\circ}$
\\$P=8.237 \pm 0.039$ yr
\\$t_0=2455522 \pm 63$ BJD
\\ \\
The resulting LTTE function is shown Fig.~\ref{fig3}. 
\\
\begin{figure}[htbp]
\centering
\includegraphics[width=13cm]{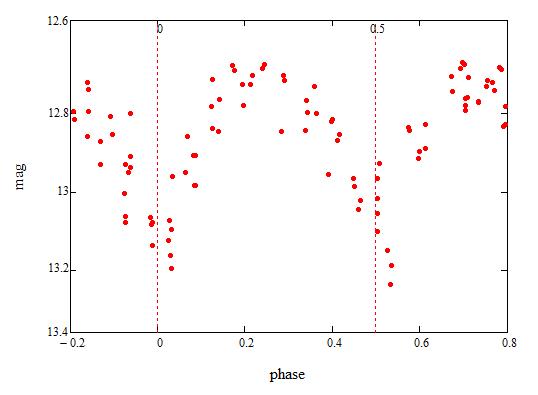}
\caption{The phase plot from the ROTSE measurements for $T_1-0.003$ day and $P_1$. }
\label{fig4}
\end{figure}

\section{ROTSE data}
Between 1989 and 2005 there are no published measurements of ToMs. However, in 1999, there are 99 photometric measurements from the Robotic Optical Transient Search Experiment (ROTSE). I make a phase plot for these data with the $T_1$, $P_1$ determined above. For the minima to be at phase 0 and phase 0.5, $T_1$ needs to be shifted by roughly -0.003 day. This is plotted Fig.~\ref{fig4}. 
\\ \\
The average time of the ROTSE observations is 2451424.26462 BJD. This is plotted on the O-C diagram  Fig.~\ref{fig3}.
\\ \\ 
My interpretation of the O-C diagram as the LTTE from a third object is in agreement with the ROTSE observations.  

\begin{figure}[htbp]
	\centering
	\includegraphics[width=13cm]{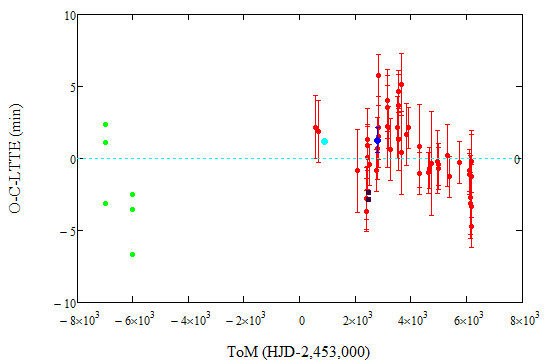}
	\caption{Green: \citet{leborgne89}; Black: \citet{hubscher11}; Blue: \citet{banfi12}; Cyan: GCVS; Red: my observations. }
	\label{fig5}
\end{figure}
\section{Conclusions}
The O-C diagram may plausibly be explained by the LTTE due to a third body, with a period of 8.2 years. CK Aqr would then be another occurrence of a W UMa star in a ternary system.
\\ \\
According to Gaia DR2, the parallax to CK Aqr is $2.4957 \pm 0.0466$ mas. The distance is then 400 pc and it may be difficult to resolve the third component. However, it may be possible to observe it spectroscopically. 
\\ \\
The residuals of the O-C data from the LTTE model are displayed Fig.~\ref{fig5}. They may seem to show a pattern instead of being random, although this is barely significant. This may come from not having enough homogeneously distributed data to make a precise fit, or because of something else; more ToM observations (over many years) would tell.
\\ \\
\setcounter{secnumdepth}{0}
\OEJVacknowledgements{The use of the on-line tool of the University of Ohio to convert HJD to BJD, at \href{http://astroutils.astronomy.ohio-state.edu/time/hjd2bjd.html}{http://astroutils.astronomy.ohio-state.edu/time/hjd2bjd.html}, is acknowledged.
\\ \\	
	The use of photometric measurements for CK Aqr from the Robotic Optical Transient Search Experiment (ROTSE), at  \href{https://skydot.lanl.gov/nsvs/nsvs.php}{https://skydot.lanl.gov/nsvs/nsvs.php},
	 is acknowledged. } \\


\begin{thebibliography}{}
	\bibliographystyle{plainnat}
	
	\bibitem[Banfi et al (2012)]{banfi12}Banfi M. et al, 2012, IBVS 6033.	
	
	\bibitem[Bilir et al (2008)]{bilir08}Bilir S. et al, 2008, MNRAS \textbf{384} 1178.
	
	\bibitem[H\"{u}bscher (2011)]{hubscher11}H\"{u}bscher J., 2011, IBVS 5984.
	
	\bibitem[Le Borgne et al (1989)]{leborgne89}Le Borgne J.F., Poretti E. and Figer A., 1989, IBVS 3316. 
	
	\bibitem[Smith et al (2002)]{smith02}Smith A.J. et al, 2002, AJ \textbf{123} 2121.



\end{thebibliography}
\end{document}